\documentclass[runningheads]{svmult}

\usepackage{makeidx}   
\usepackage{graphicx}  
\usepackage{subeqnar}  
\usepackage{multicol}  
\usepackage{physprbb}  
\makeindex             

\begin{document}
\title*{ R=100,000 Mid-IR Spectroscopy of UCHII Regions: High Resolution is Worth it!}
\toctitle{R=100,000 Mid-IR Spectroscopy of UCHII Regions: 
\protect\newline High Resolution is Worth it!}
%
%
\titlerunning{Mid-IR Spectroscopy of UCHII Regions}
%
\author{D.T. Jaffe\inst{1}
\and Q. Zhu\inst{1}
\and J.H. Lacy\inst{1}
\and M.J. Richter\inst{2}
\and T.K. Greathouse\inst{1}}
\authorrunning{D.T. Jaffe et al.}
%
%
\institute{Department of Astronomy, University of Texas, Austin TX
\and Department of Physics, University of California, Davis CA}

\maketitle              

\begin{abstract}
Ultracompact HII regions are signposts
of massive star formation and their properties provide diagnostics for
the characteristics of very young O stars embedded in molecular
clouds.  While radio observations have given us a good picture
of the morphology of these regions, they have not provided
clear information about the kinematics.  Using high spectral
resolution observations of the $12.8\mu$m [NeII] line, it 
has been possible for the first time to trace the internal kinematics
of several ultracompact HII regions.  We find that the
motions in the cometary ultracompact HII regions
MonR2 and G29.96-0.02 are highly organized.  The
velocity patterns are consistent with parabolic
ionized flows along a neutral boundary layer.
\end{abstract}

\section{Introduction}
An understanding of the formation of massive stars and of the
effects of these stars on their environments is critical to an understanding
of star formation overall.  Most stars of all masses form in
clusters containing massive stars \cite{ladalada}. High mass
stars and the regions of the ISM that they excite are what we
use to trace star formation in other galaxies. The formation
of massive stars ultimately destroys the star forming clouds.

Ultracompact HII (UCHII) regions are one of the earliest
manifestations of massive stars. High extinction and substantial
emission by overlying dust make it impossible to detect the 
photospheres of very young O stars directly at any wavelength.
The UCHII regions, however, become detectable
in the radio continuum and in mid-IR fine-structure lines 
while young O stars are still embedded 
in their natal envelopes, as soon as the stars have begun to emit
any significant amount of Lyman continuum radiation.

We have been using the $12.8\mu$m fine-structure line of [NeII]
to study the morphology and kinematics of a sample of UCHII
regions.  Our ultimate goal is to understand the natures of the
regions and how they depend on the age and mass of the exciting
star, as well as on the properties of any disk or envelope.  We
would also like to learn how long the UCHII regions last and
how the evolution of these regions affects the surrounding cloud.
In this paper, we summarize results from our initial studies of
UCHII regions with an apparently cometary morphology \cite{jaffe}
\cite{zhu}.

%

\section{Observations}

Radio recombination line observations, even when made with aperture
synthesis instruments with high spatial resolution, have not been
able to reveal the kinematics of UCHII regions in detail.
In at least some UCHII regions, radio observers have been able to
find evidence for a mix of broad and narrow hydrogen recombination
line components \cite{depree},
for systematic velocity gradients \cite{garay} \cite{tieftrunk}, 
and for multiple velocity
components.  In this section, we describe the observations we have made
with the TEXES spectrometer \cite{lacy} at the IRTF and the advantages
these observations have over radio recombination line results.

Over the past few years, we have mapped a small sample of UCHII regions
using the mid-IR spectrometer, TEXES on the NASA IRTF in the $12.8\mu$m
fine structure line of [NeII].  TEXES
has sufficiently good spectral resolution (3.4 km s$^{-1}$ at $12.8\mu$m)
to be able to resolve structure in the shapes of the nebular emission
lines.  With a 1.4$^{\prime\prime}$ slit,
the TEXES-IRTF combination has a 
spatial resolution that compares favorably
with that of most existing VLA maps of radio recombination lines.
We have mapped our sample of UCHII regions by scanning the TEXES slit
(while holding the telescope secondary fixed) across the sources and
adjacent blank sky.  At the high spectral resolution of TEXES, it is
then possible to produce accurate maps of the [NeII] intensity 
distribution by subtracting the average off-source spectrum from 
the spectrum at each point where emission may be present.

Two of the main advantages of observing UCHII regions in the mid-IR 
derive from properties of the [NeII] line rather than of the
instrument.  The first of these is that the line is incredibly bright.
Our observations of Mon R2, show a ratio of [NeII] line flux to 5 GHz radio 
continuum flux density of 1.5$\times 10^{-13}$ Wm$^{-2}$Jy$^{-1}$ 
making it possible to
\begin{figure}[t]
\begin{center}
\includegraphics[width=.7\textwidth]{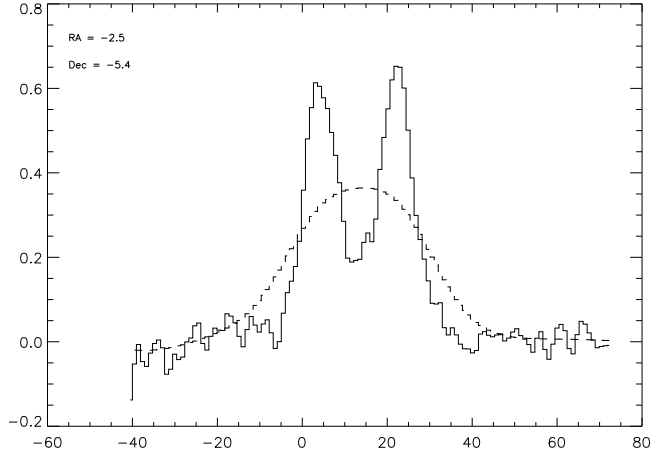}
\end{center}
\caption[]{[NeII] spectrum toward the interior of MonR2.  The solid line shows
the spectrum at full resolution($\Delta$V= 3.4 km s$^{-1}$).  
The dashed line shows the
same spectrum smoothed with a 20 km s$^{-1}$ Gaussian (the thermal broadening
of hydrogen lines at 10$^4$ K, or the resolution of a resolving power 15,000
spectrometer).}
\label{hvsne}
\end{figure}
obtain S/N$\sim$50 spectra of narrow lines with TEXES
in only 10 seconds of integration in regions with radio continuum
flux densities of only 10 mJy per beam \cite{jaffe}.  The other advantage 
derives from neon's large atomic weight.  In gas where
the motions are entirely thermal, the neon line will be only 22\% the
width of a hydrogen recombination line.  The much smaller thermal widths
allow neon spectra, when observed at the high resolution possible with
TEXES, to reveal mass motions at much lower levels than
is possible with hydrogen line observations.  Figure~\ref{hvsne} shows a [NeII]
spectrum toward one position in the MonR2 UCHII region\cite{jaffe}. 
Superposed on the observed
spectrum is a model showing what an observation of the same region
in a hydrogen recombination line would have looked like.
The thermal broadening of the hydrogen line completely masks the complex
structure apparent in the [NeII] observations.

\section{Progress Report}

In our first paper on TEXES mapping of [NeII] emission from UCHII regions,
we described observations of Mon R2 \cite{jaffe}.  
The two remarkable results in these
measurements were (1) that the lines were locally quite narrow (as small
as 8 km s$^{-1}$ FWHM) and (2) the velocity structure of the region was 
predominantly in the form of large-scale and well-organized features.  
These features did not, however, fit the simple pattern of an outflow
or an expanding shell.

Both Mon R2 and the second region we have analyzed in detail, G29.96-0.02
\cite{zhu}, have a cometary morphology, a characteristic they share with
roughly 20\% of all UCHII regions \cite{wood}. Over the past 15 years,
a number of authors have modeled cometary UCHII regions as flowing
ionized gas behind
a bow shock.  In the models, the ionizing O star has a peculiar velocity
with respect to its parent cloud.  The bow shock forms
at the interface of the high velocity
stellar wind and the dense molecular gas flowing toward the star. Numerical
simulations have been compared in detail to the radio continuum morphologies
of cometary UCHII regions \cite{maclow}.  More recently, an analytic model, 
including at least some of the relevant physics has been developed 
\cite{wilkin}.

\begin{figure}[t]
\begin{center}
\includegraphics[width=.7\textwidth]{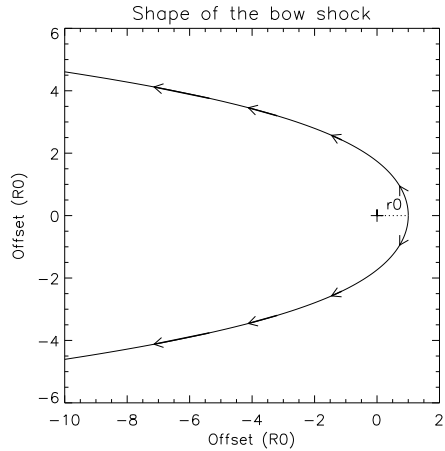}
\end{center}
\caption[]{Calculated shape and velocity field of a bow shock for the
model that best fits G29.96-0.02 \cite{zhu}.  The cross at (0,0) shows
the position of the exciting star and the arrows show the velocity 
of the flowing gas in the frame of the star.}
\label{parabola}
\end{figure}

The more detailed kinematic information made available by the high 
signal to noise and low intrinsic line widths of the [NeII] line makes
possible a more detailed comparison of UCHII region motions to the parabolic
flows predicted by bow shock models.
In Zhu et al. \cite{zhu}, we compare [NeII] mapping of both 
Mon R2 and G29.96-0.02 to such a parabolic flow picture.  
We have developed a simple numerical model to predict the kinematics in
cometary UCHII regions resulting from the interaction of the stellar
wind and the inflowing molecular gas in the bow shock.  The model 
accounts for the ionization equilibrium, momentum flow, and pressure
differentials in the ionized parabolic shell \cite{zhu}.
Figure~\ref{parabola} shows how the gas flows along the shell that
forms at the boundary between the fast stellar wind and the moving
(in the frame of the star) molecular gas.

A comparison of models and observations
shows that the fairly simple parabolic flow pattern produced by the
bow shock models describes the UCHII 
regions very well.  Zhu et al. \cite{zhu} will discuss the issues 
involved in the physics of the regions.  Here we simply point out the
observational fact of the remarkably good agreement between the
data and the models.  

\begin{figure}[b]
\begin{center}
\includegraphics[width=.7\textwidth]{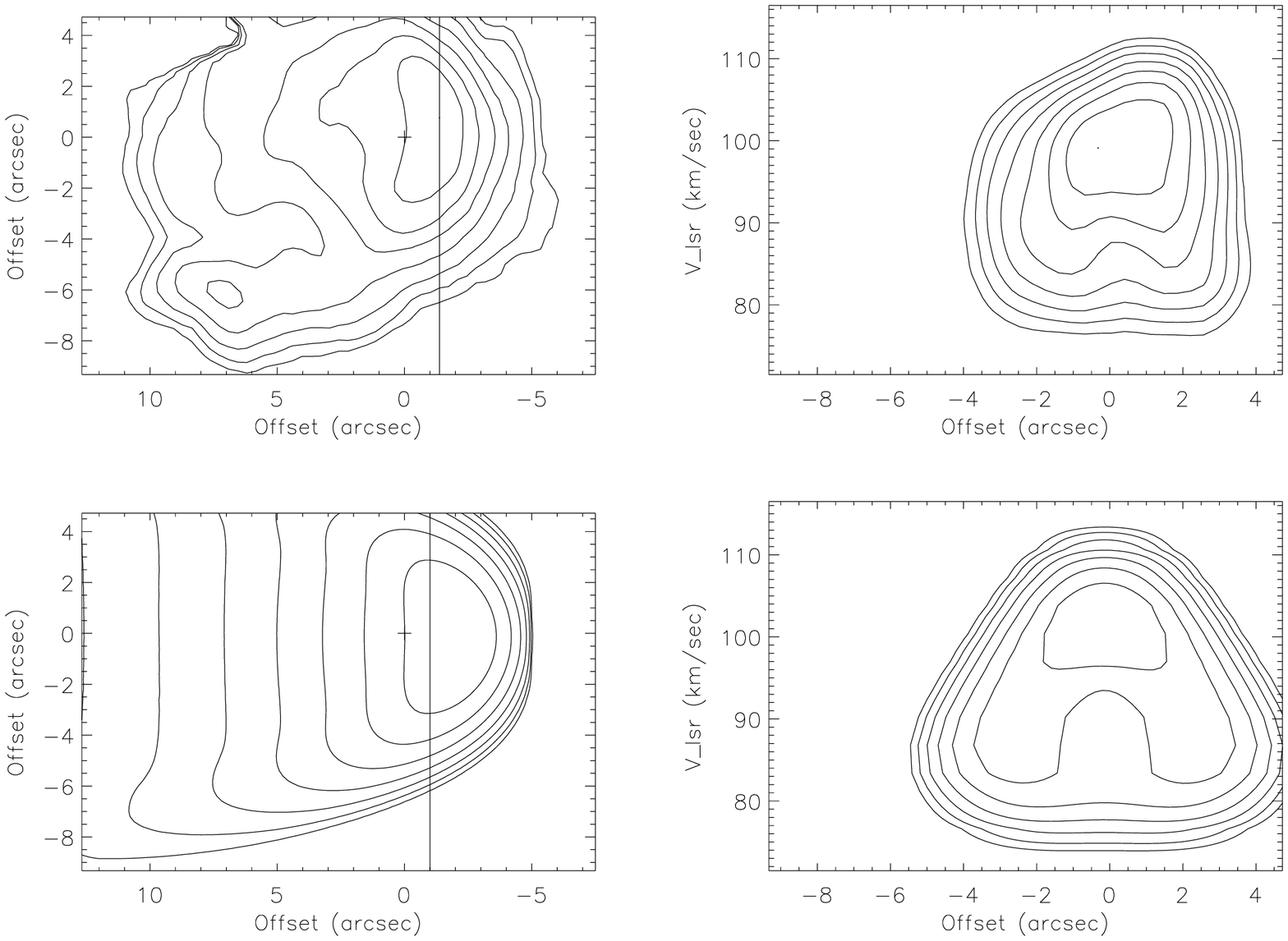}
\includegraphics[width=.7\textwidth]{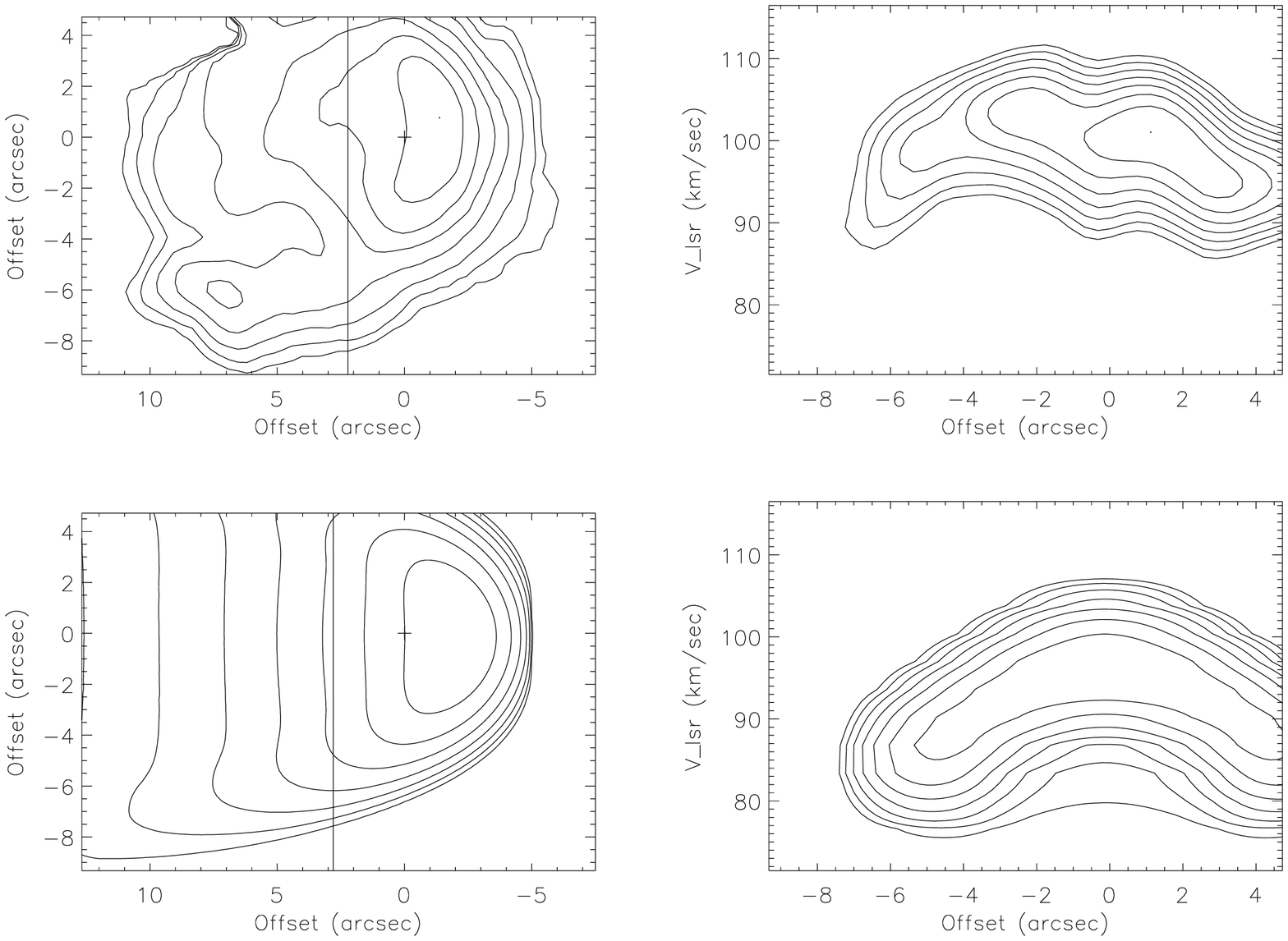}
\includegraphics[width=.7\textwidth]{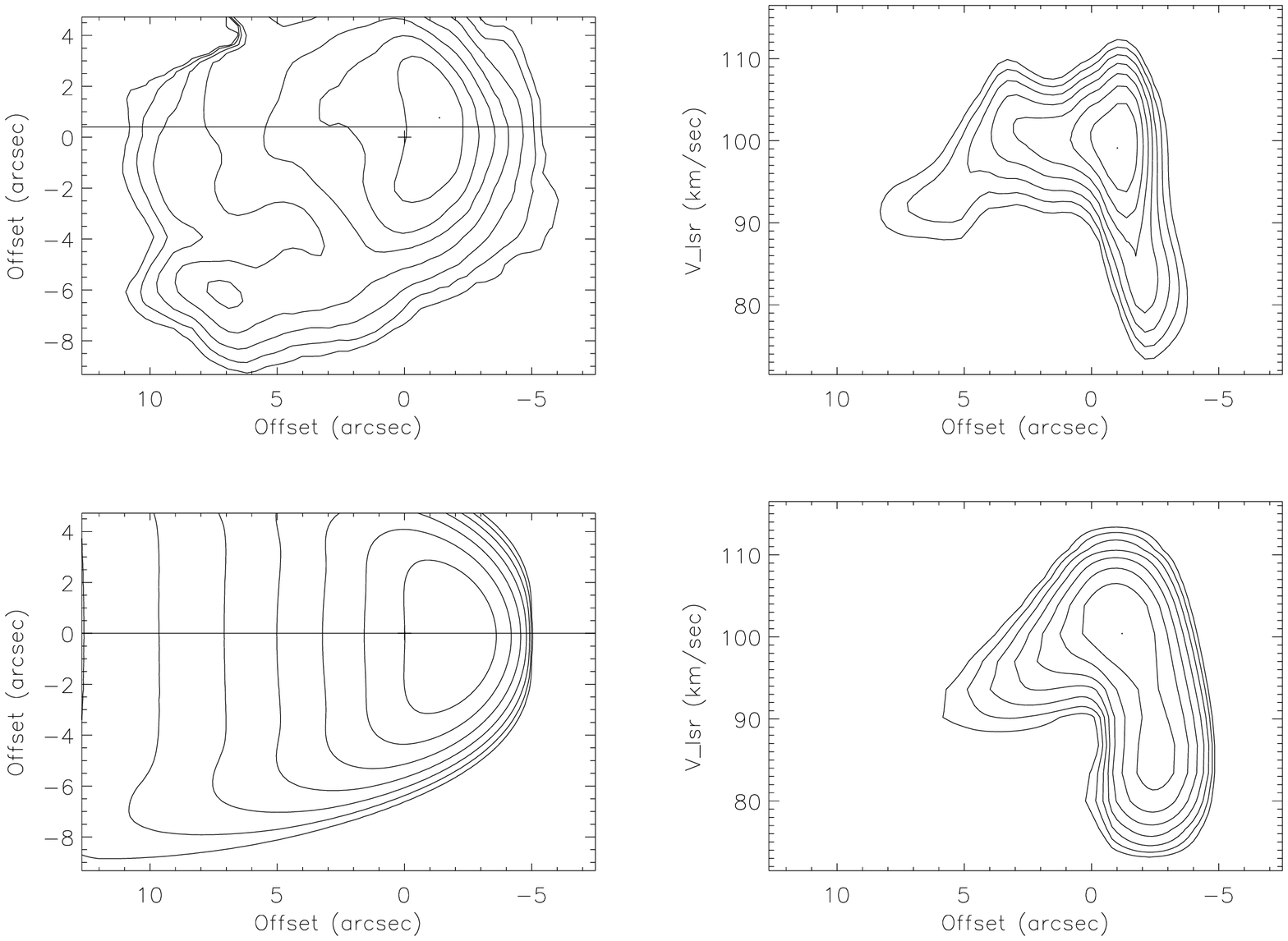}
\end{center}
\caption[]{The left column shows alternately [NeII] integrated
  linestrength maps toward G29.96-0.02 
  \cite{zhu} and integrated 
  line representations of a bow shock model for this source, in
  each case with a superposed line showing the location of the
  position-velocity diagram shown to the right of each map. For
  all three p-v cuts, the observed and model diagrams are 
  remarkably similar.}
\label{g29}
\end{figure}

Figure \ref{g29}
compares, in succeeding panels, position-velocity diagrams for
a parabolic flow model and p-v diagrams of G29.96-0.02 itself.  Cuts both
parallel and perpendicular to the symmetry axis show remarkable
similarity when one compares the data and the model at the 
corresponding points within the structure.


\section{Future Directions}

While the close match of the bow shock models and the observed kinematics
of Mon R2 and G29.96-0.02 is extremely encouraging, it marks more of a
beginning than an end to our investigation of the nature of UCHII regions.
The close match between models and observations says that the kinematic
part of the models is likely correct; any picture of cometary 
HII regions must
produce a systematic flow along a relatively thin, approximately 
paraboloidal shell.  The high frequency of cometary UCHII regions and the 
large peculiar stellar velocities we need in our fits, however, argue
that we have not yet produced a complete description of the physics of
the fluid flow.  Zhu et al. \cite{zhu} discuss some of the issues 
involved and our group continues work on more elaborate hydrodynamical
models.
The bow shock picture, in any case, can explain neither all of the 
features observed in the two sources discussed here (in particular,
the compact, jet-like feature in Mon R2 \cite{jaffe}) nor the
general kinematics of sources with non-cometary morphologies.
A combination of more detailed hydrodynamic modeling and the
higher spatial resolution afforded by 8m class infrared telescopes
will make it possible for us to get at the causes of the large-scale
motions in these other classes of sources.

The authors were visiting astronomers at the Infrared Telescope 
Facility, which is operated by the University of Hawaii under Cooperative
Agreement no. NCC 5-538 with the National Aeronautics and Space
Administration, Office of Space Science, Planetary Astronomy Program.
This work was supported in part by National Science Foundation Grant
AST-0205518 to the University of Texas at Austin.

%


\begin{thebibliography}{8.}
\addcontentsline{toc}{section}{References}

\bibitem{depree} De Pree, C.G., Wilner, D.J., Mercer, A.R., Davis, L.E., Goss, W.M. \& Kurtz, S.: ApJ, 600, 286 (2004)

\bibitem{garay} Garay, G., Rodriguez, L.F., \& van Gorkom, J.H.: ApJ, 309, 553 (1986)

\bibitem{jaffe} Jaffe, D.T., Zhu, Q., Richter, M.J., \& Lacy, J.H: ApJ 596, 1053 (2003)

\bibitem{lacy} Lacy, J.H., Richter, M.J., Greathouse, T.K., Jaffe, D.T., \&
  Zhu, Q.: PASP, 114, 153 (2002)

\bibitem{ladalada} Lada,C. \& Lada, E.A.: ARAA, 66 (2003)

\bibitem{maclow} Mac Low, M.M., van Buren, D., \& Churchwell, E.: ApJ
    369, 395 (1991)

\bibitem{tieftrunk} Tieftrunk, A.R., Gaume, R.A., Claussen, M.J., Wilson, T.L.,
  \& Johnston, K.J., A\&A, 318, 931 (1997)

\bibitem{wilkin} Wilkin, F.P.: ApJ, 459, L31 (1996)

\bibitem{wood} Wood, D.O.S., \&  Churchwell E.: ApJS 69, 831 (1989)

\bibitem{zhu} Zhu, Q., Lacy, J.H., Jaffe, D.T., Greathouse, T., \&
  Richter, M.J.: in preparation (2004)

\end{thebibliography}
\end{document}